\documentclass[prl,twocolumn,graphicx,amssymb,floatfix]{revtex4-1}
\usepackage{graphicx}
\usepackage{color}
\usepackage{soul}
\usepackage{float}
\begin{document}

\title{An Improved Experiment to Determine the `Past of a Particle' in the Nested Mach-Zehnder Interferometer}
\author{Alon Ben-Israel$^1$,  Lukas Knips$^{2,3}$, Jan Dziewior$^{2,3}$, Jasmin Meinecke$^{2,3}$, Ariel Danan$^1$, Harald Weinfurter$^{2,3}$, and Lev Vaidman$^1$}
\affiliation{  $^1$ Raymond and Beverly Sackler School of Physics and Astronomy,
 Tel-Aviv University, Tel-Aviv 69978, Israel\\$^2$ Max-Planck-Institut f\"{u}r Quantenoptik, Hans-Kopfermann-Stra{\ss}e 1, 85748 Garching, Germany\\
$^3$ Department f\"{u}r Physik, Ludwig-Maximilians-Universit\"{a}t, 80797 M\"{u}nchen, Germany
}

\begin{abstract}
We argue that the modification proposed by Li \textit{et al}. [Chin. Phys. Lett. {\bf 32}, 050303 (2015)] to the experiment of Danan {\it et al.} [Phys. Rev. Lett. {\bf 111}, 240402 (2013)] does not test the past of the photon as characterised by local weak traces. Instead of answering the questions: (i) \textit{Were the photons in $A$?} (ii) \textit{Were the photons in $B$?} (iii) \textit{Were the photons in $C$?} the proposed experiment measures a degenerate operator answering the questions: (i) \textit{Were the photons in $A$?} (ii) \textit{Were the photons in $B$ and $C$ together?}
A negative answer to the last question does not tell us if photons were present in $B$ or $C$.
A simple variation of the modified experiment does provide good evidence for the past of the photon in agreement with the results Danan {\it et al.} obtained.
\end{abstract}
\maketitle

Li \textit{et al}. \cite{china_experiment} recently proposed an `ideal' experiment designed to determine the past of a particle passing through the nested interferometer analyzed by Danan \textit{et al}.   \cite{asking_photons}.
They proposed to use an alternative method for observing the location of the photon based on Kerr media to challenge and refute Danan's claim that the past of a photon in this interferometer is described by disconnected paths.

In this Letter we  analyze the method of Li \textit{et al}. and find that their proposed experiment is not a good test of the past of the photon.
However, a modification of their experiment does provide a correct alternative measurement of the past of the photon, which, as we believe, will reveal the disconnected paths that Danan \textit{et al}. have characterized.

First, we ask in what way the proposed experiment is `ideal'.
In standard quantum mechanics there is no concept of the particle path or the past of a particle.
The past of a particle is not defined, and so there cannot be an `ideal' way to find it.
The approach which does not allow to talk about particles at intermediate times between measurements saves us from having to consider seemingly paradoxical results, but at the same time limits the possible insight we may gain by considering this concept.

Several approaches have been suggested that allow us to discuss the past of particles in quantum mechanics, associating trajectories to each particle.
One of those is the de Broglie-Bohm interpretation of QM, in which the trajectories of particles are determined by the wavefunction via a guiding equation \cite{DBB}.
If the wavefunction of the particle is a well-localized wave packet, the Bohmian trajectory of the particle coincides with the trajectory of the wave packet.
For an evolving wave packet that splits into several wave packets, of which only one reaches the final destination via a continuous path, the trajectory of this packet can be defined as the path of the particle.
This is the `common sense' approach advocated by Wheeler \cite{wheeler_commonsense}: the particle went through this path because it could not come through any other path.
Recently, Vaidman  \cite{past_quantum_particle} proposed another definition: the past of the particle is described by the locations where a particle leaves a weak trace.
Danan's experiment was designed to measure this weak trace.

The measurement of the trace in the experiment of Danan \textit{et al}. invariably spoils the perfect interference of the inner interferometer and creates some leakage in its dark port. Apparently, this leakage is what made the original experiment `not ideal' in the eyes of Li \textit{et al}. This view is supported by the fact that the leakage is crucial for explaining the results of Danan \textit{et al}.: the meter of their experiment was a transversal degree of freedom of the photon itself.
The trace, `written' on the wave function of the photon, could not be observed by the quad-cell detector placed outside the interferometer without the leakage towards it.
From this perspective the proposal of Li \textit{et al}. to place the meter inside the interferometer is a desired change. The trace is recorded where it is created.  Therefore, we need not to confront the question: How does the external detector get the information about the trace inside the inner interferometer if only a tiny leakage passes from the place with the trace toward the detector?

However, the conduction of a measurement which detects the weak trace of the photon inside the interferometer without testing the traces in each of its arms separately,
is a step in the wrong direction. 
The setup with the nested interferometers is analogous to a {\it three box paradox} \cite{AV91}, where the paths of the interferometer correspond to the {\it three boxes}.
We know that if we look in arm $A$ we find the photon with certainty and also, if we look at arm $C$ instead, we find it there with certainty too. But if we test the presence  of the photon  anywhere in $B$ or $C$ without resolving these two paths, we are certain not to find it, since it is equivalent to testing its presence in $A$. It has been proven \cite{AV91} that if a usual (strong) measurement of an observable performed on pre- and post-selected system yields a particular eigenvalue with certainty, a weak measurement of this observable must yield the same value. The experiment of Li \textit{et al}. is such a weak measurement of the projection onto $B$ and $C$ together, so it must yield null result.

The outcomes of weak measurements are weak values, and the experiment can be understood also in this language. In the three-box setup, the weak values of the projection operators on different boxes are:
\begin{equation}
({\rm \bf P}_C)_w=1,~~~({\rm \bf P}_A)_w=1,~~~({\rm \bf P}_B)_w=-1.  \end{equation}
  The weak values are additive, so
\begin{equation}
\left({\rm \bf P}_B + {\rm \bf P}_C \right)_w=\left({\rm \bf P}_B  \right)_w+\left( {\rm \bf P}_C \right)_w
=-1+1=0.   \end{equation}

Vaidman's principle is that the pre- and post-selected photon was in every place where it left a local trace. Any nonvanishing weak value of a local operator in a particular place leads to a local weak trace.
Li \textit{et al}.'s experiment does not observe all these local traces.
It weakly measures the projection onto $B$ and $C$ together.

\begin{figure}[b]
  \centering
    \includegraphics[width=0.4\textwidth]{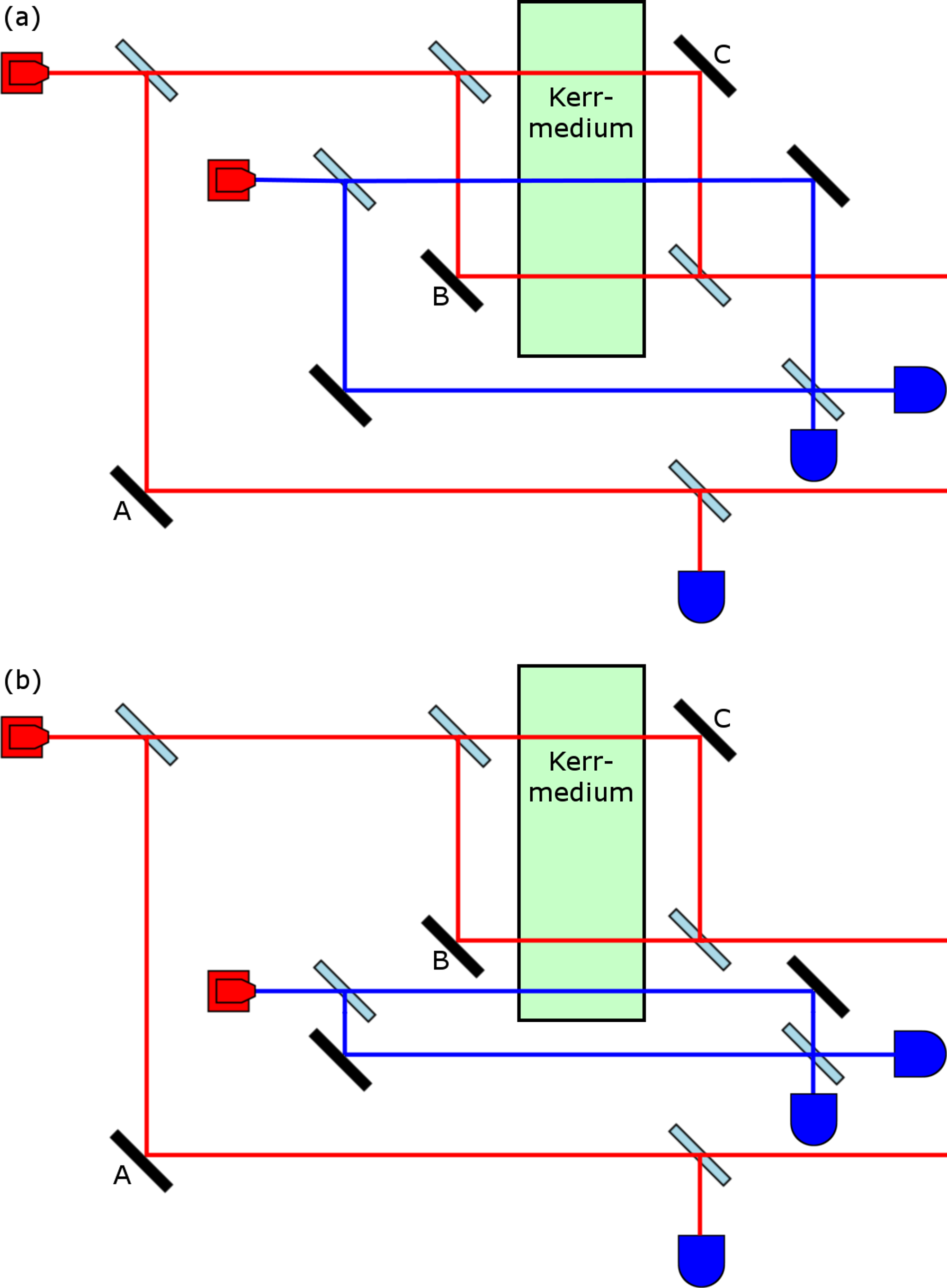}
  \caption{(a) The experimental setup proposed by Li \textit{et al}. in which the probe photon's path runs through the middle of the Kerr media. (b) Our proposed modification, in which the path of the probe photon is moved to a region in the Kerr media in close proximity of the arm of the interferometer wherein the presence of the photon is tested.}
  \label{weak_angle_fig}
\end{figure}

Even though
according to the definition proposed by Vaidman
the photon was in $B$, and also was in $C$, the influences of the photon in the two places  on the meter of Li \textit{et al}.
cancel each other.
The meter in their experiment is the phase acquired by the probe photon passing in the Kerr media in the middle of the inner interferometer, see Fig. \ref{weak_angle_fig}a.
The photon influences the probe photon due to its presence   in both arms, $B$ and  $C$, but the influences are in opposite directions  resulting in the null outcome.
This is possible because contrary to the case of a photon that is pre-selected only in a superposition of being in different arms of the interferometer causing a mixture of evolutions of the probe photon, the  pre- and post-selected photon yields a superposition of the evolutions of the probe photon \cite{Aharonov1990} which can cancel each other.

A small modification of the proposed experiment {\it is} suitable for measuring the local trace inside the interferometer.
We just have to move the path of the probe photon near the place where we want to observe the trace, see Fig. \ref{weak_angle_fig}b.
Repeating the experiment with a probe photon passing in different regions inside the nested interferometer (or adding more photon-meter interferometers) will provide the full information about the past of the photon.
These local measurements will necessarily destroy the perfect interference of the inner interferometer leading to some unavoidable leakage.
However, the weak trace left by this leakage is vanishingly small.
Indeed, an identical coupling in all arms of the interferometer which causes the traces of order $\epsilon$ in arms $A$, $B$, and $C$ will lead to the trace in the dark port proportional to
 $\epsilon^2$.
In the weak limit of $\epsilon\rightarrow0$ the ratio of the magnitudes  of these traces goes to zero and the trace proportional to $\epsilon^2$ can be neglected.
In this sense, the photons are present (leave a trace) in the arms $B$ and $C$ inside the inner interferometer, but not in the arms leading in and out of it.
For more discussion, see \cite{PF, comment_potocek, Salih, comment_salih}.

The modified proposal of Li \textit{et al}. is conceptually a better experiment for observing the past of a photon defined as the regions where it leaves a weak trace.
It is a direct measurement with an external device.
Moreover, it is a genuinely quantum experiment since its results cannot be explained by Maxwell's equations of the electromagnetic field of the laser, as they were explained by Danan \textit{et al}. in their experiment.
However, it is much more challenging.
In view of a recent proposal \cite{Steinberg}, it is on the verge of technological feasibility.
Still, the experiment of Danan \textit{et al}., even if it has an alternative explanation, is a good demonstration of the past of a pre- and post-selected photon.

In conclusion, the null result claimed by Li \textit{et al}. is obtained not because there was no effect, but because in their measurement the effects of the photon on the meter interferometer from the arms $B$ and $C$ of the inner interferometer cancel each other.
Shifting the path of the meter interferometer from the center of the inner interferometer would reveal the weak trace of the photon there.
Such a modified experiment will be an improvement over the experiment by Danan \textit{et al}., which is  worth performing.

This work has been supported in part by the German-Israeli Foundation for Scientific Research and Development Grant No. I-1275-303.14.


\end{document}